\documentstyle[12pt, times]{article}

\newcommand{\be}{\begin{equation}}
\newcommand{\ee}{\end{equation}}

\newcommand{\bary}{\begin{eqnarray}}
\newcommand{\eary}{\end{eqnarray}}

\def \mwt {{\mid {{w_t+k}\over {w_t-k}}\mid}}
\def \mwl {{\mid {{w_l+k}\over {w_l-k}}\mid}}
\def \mnt {{\mid {{1+n_t}\over {1-n_t}}\mid}}
\def \mnl {{\mid {{1+n_l}\over {1-n_l}}\mid}}
\begin{document}
\title{Cerenkov radiation of longitudinal photons by neutrinos} 
{\author{Sarira Sahu$^{1}$\\          
Theory Group, Physical Research Laboratory,\\
Navrangpura, Ahmedabad-380 009\\ India}
\date{ }
\maketitle
\footnotetext[1]{email:sarira@prl.ernet.in}
\thispagestyle{empty}
\begin{abstract}

In a relativistic plasma neutrino can emit plasmons by the Cerenkov
process which is kinematically allowed for a range of frequencies
for which refractive index is greater than one. 
We have calculated the rate of energy emission by this process.
We compute the energy deposited in a stalled supernova shock wave
by the Cerenkov process and find that it is much smaller than the 
Bethe-Wilson mechanism.
\end{abstract}
\vfill
\eject

\section{Introduction}

The electromagnetic wave passing throuth the plasma, 
is modified because of the mobile charged particle in the medium and
it consists of coherent vibration of the electromagnetic field
as well as the density of the charged particles\cite{semikoz}.
Photons no longer propagate at the speed of light and satisfy the
dispersion relations for transverse and
longitudinal modes. Transverse photons are similar to the ordinary photons
in the vacuum and longitudinal photons are the collective excitation of
the plasma known as "plasmon". 
Cerenkov radiation is emitted when a charge particle moves through a
medium with a velocity greater than $c/n$, $n$
being the refractive index of the 
medium. This is also true for neutral particles with non-zero magnetic and
or electric dipole moments. 
For Cerenkov radiation to take place in the medium,
the refractive index of photon should satisfy the condition 
$n= |{\bf k}|/w~ >~ 1$, where $\bf k$ and $\omega$ are momentum and frequency
of the emitted photon respectively.
Recently several authors have considered the
Cerenkov  radiation emitted by neutrinos as 
they pass through a medium\cite{pal1,altherr,grimus,suzuki,raffelt}. 
Olivo, Nieves and Pal have recently shown that, neutrino can emit Cerenkov
radiation even in the massless limit and having no  
electromagnetic dipole moments\cite{pal1}. 

In a relativistic plasma neutrinos can loose energy by the Cerenkov radiation
of plasmon. This takes place: firstly because neutrinos acquire an effective
charge  in the medium, 
by coupling to the electromagnetic
field through electrons and positrons 
in the plasma\cite{pal1,altherr,notzold}, 
which is shown in the Feynman diagrams in 
figure 1. 
Secondly there is a range of 
frequencies of plasmon for which the refractive index $n~>~1$, and the 
Cerenkov process is kinematically allowed. We compute the rate of energy
radiated by neutrinos (even those with zero electromagnetic dipole moments)
in relativistic plasma. We find that the rate of energy radiated in the 
form of plasmon is 
$$S\simeq {T\over {8\pi^2\alpha}} G^2_F C^2_V P_{l}^2 (n-1)^3 E_1^2, $$
where $E_1$ is the incoming neutrino energy. 
We calculate the energy deposition by this process in the stalled 
shock wave of the supernova and compare with the Bethe-Wilson (BW) 
mechanism of shock revival. We found that the Cerenkov process is
very weak compared to the BW mechanism.

\section{Cerenkov process}

The dispersion relations satisfy by the transverse and longitudinal
modes of photon depend on the properties of the plasma. 
In the relativistic limit the dispersion relations are given by\cite{braaten} 
\be
w_t^2 =k^2 + w_p^2 {3 w_t^2\over {2 k^2}}
\Big (1- {(w_t^2 - k^2) w_t\over {w_t^2 2 k}} 
log\mwt \Big )~~~~~0\le k <\infty
\ee
and
\be
w_l^2 = w_p^2 {3 w_l^2\over k^2} \Big ( {w_l\over {2 k}} 
log\mwl - 1\Big )~~0\le k <\infty.
\ee
for the transverse and longitudinal modes respectively.
In the above equations $w_p$ is the plasma frequency and $k$ is the photon 
momentum.
In the medium with refractive index $n_{\alpha}={|k|/ w_{\alpha}}$, the above
dispersion relations can be expressed as
\be
({w_t\over w_p})^2 = {3\over {2 n^2_t (1-n^2_t)}} 
\Big (1 - {(1-n^2_t)\over {2 n_t}} 
log\mnt\Big )
\ee
and
\be
({w_l\over w_p})^2 = 
{{3\over n^3_l}\Big ({1\over 2} log\mnl -n_l \Big )}.
\label{wl}
\ee
For $n > 1$, $({w_t/ w_p})^2$ is always negative but 
for a range of $n > 1$;
$({w_l/ w_p})^2$ is positive which is shown in figure 1.
From the dispersion relations we see that the refractive index $n_t=w_t/|k|$
of transverse photon is always less than one, so transverse photon can not
be emitted by Cerenkov process in a plasma. On the other hand, 
for longitudinal photon we see that
$n_l$ can be greater than one. So there can be plasmon emission 
by Cerenkov process. 
Recently several authors have considered the Cerenkov radiation by 
neutrinos in a medium\cite{pal1,grimus}. 
Neutrino properties get modified 
when propagates through a medium as a consequence of the weak interaction 
with the background particles\cite{notzold}.
It has been shown earlier 

\newpage
\vspace*{4.5in}
\begin{figure}
\includegraphics{figure1.ps}
\end{figure}
\vspace*{0.8in}
\noindent {~~~~~~~~~~~~~~~Fig. 1: RHS of eq.(\ref{wl}) is ploted 
as a function of $n_l$. }
\vskip .5cm

\noindent
that neutrino
acquires an effective charge in the medium by coupling to the electromagnetic
field through electrons and positrons in the plasma\cite{pal1,altherr}.

Here we will consider the Cerenkov  process
\be
\nu(p_1) \rightarrow \nu(p_2) + \gamma(k)
\ee
in the medium,
where $\gamma(k)$ is the plasmon (longitudinal photon) emitted with a
momentum k. Feynman diagram for the above processes are shown in figure 2.

\newpage
\begin{figure}
\vskip 5cm
\includegraphics{w.ps}
\includegraphics{z.ps}
\end{figure}
\vskip -2cm
\noindent{~~~~~~~~~~Fig. 2: Feynman diagram for 
neutrino photon coupling through $W$ and $Z$
~~~~~~~~~~~~~~~~~~~~exchange in the medium.}
\vskip .5cm

\noindent The matrix element for the above process is given by
\be
{\cal M} = {G_F\over \sqrt{2}} 
\Gamma^{\alpha\mu}\epsilon_{\mu}({\bf k},\lambda)
{\bar u}(p_2)\gamma_{\alpha} (1-\gamma_5) u(p_1).
\label{mm}
\ee
where $\Gamma^{\alpha\mu}(w,{\bf k})$ is the 
effective vertex for the plasmon 
interac ing with the neutrino current. This effective vertex 
is due to the 
W and Z in the loops in the Feynman diagram\cite{altherr,braaten}. 
The vertex tensor is gauge invariant quantity, as
$k_{\mu}\Gamma^{\alpha\mu} = 0$\cite{altherr,braaten}. 
The effective vertex tensor is\cite{braaten}
\bary
\Gamma^{\alpha\mu}(w, {\bf k})&=&{1\over {\sqrt{4\pi\alpha}}}
\Big (
C_V P_{l}
(1, {w\over k}{\hat{k}})^{\alpha} (1, {w\over k}{\hat {k}})^{\mu} 
\nonumber\\
&+& g^{\alpha i}
\Big [C_V P_t \Big (\delta^{ij}-{\hat k^i}{\hat k^j}\Big )
+C_A\Pi_A \Big (i\epsilon^{ijk}\Big )
\Big ]g^{j\mu}
\Big )
\label{vx}
\eary
where $C_V$ and $C_A$ are vector and axial vector coefficients,
$\epsilon_{\mu}(k, \lambda)$ is the 
polarization vector, $u(p_i)$ is
the neutrino spinor and $\alpha$ in the denominator is the
electromagnetic coupling constant. The functions $P_{l}$, $P_t$ and $\Pi_A$
are longitudinal, transverse and axial polarization functions 
respectively\cite{braaten}.
As we have already shown, the longitudinal part will only contribute to the 
above process, so transverse and axial parts (second and third terms in
eq.(\ref{vx}))
of the vertex are ignored. The
longitudinal polarization function for plasmon is given by\cite{braaten}
\be
P_{l}=3 w_p^2 \Big ({1\over {2 n_l}} log\mnl -1 \Big ),
 \ee
where $w^2_p = {4\pi e^2 N_e/ m_e} $ is the plasma frequency.
From eq.(\ref{mm}),  $|{\cal M}|^2$ is 
\be
|{\cal M}|^2 = {G_F^2\over 2} 
\sum_{\lambda} \epsilon_{\mu}({\bf k}, \lambda) \epsilon^{*}_{\delta}(
{\bf k}, \lambda)
\Gamma^{\alpha\mu}\Gamma^{*\beta\delta}
8\Big (p_{2\alpha}p_{1\beta} - (p_1.p_2) g_{\alpha\beta} + p_{2\beta}p_{1\alpha}
\Big ).
\label{m2}
\ee
Henceforth we will be using $n_l=n$ and $w_l = w$.
Using the polarization sum in the medium
\be
\sum_{\lambda} \epsilon_{\mu}({\bf k}, {\lambda})
\epsilon^*_{\delta} ({\bf k}, {\lambda})=
-g_{\mu\delta} + (1-{1\over n^2}) W_{\mu} W_{\delta} +
{1\over {n^2 w}} (W_{\mu}k_{\delta}  + k_{\mu} W_{\delta})
-{1\over {n^2 w^2}} k_{\mu}k_{\delta},
\ee
with $W_{\mu}=(1, 0)$ the center of mass velocity of the medium.
Putting this in eq.(\ref{m2}) we obtain for the longitudinal part,
\bary
|{\cal M}|^2 &=& {G_F^2\over 2} C_V^2 P_{l}^2 {1\over {4\pi\alpha}}
8 (1-{1\over n^2})\Big [2 (E_2 - ({\bf p_2.k}) {w\over k^2}) 
\nonumber\\
& &(E_1 -({\bf p_1.k}){w\over k^2})
-(E_1 E_2 - ({\bf p_1.p_2}) (1-{w^2\over k^2})\Big ],
\eary
where $p_1=(E_1, {\bf p_1})$, 
$p_2=(E_2, {\bf p_2})$ and $k=(w, {\bf k})$ are the four-momenta of 
incoming neutrino, outgoing neutrino and outgoing photon respectively.
Then the total energy emitted from a single process is
\be
S={1\over {2 E_1}}\int {d^3p_2\over {2 E_2 (2\pi)^3}}
{d^3k\over {2 w (2\pi)^3}} w (2\pi)^4 \delta^4 (p_1 - p_2-k) |{\cal M}|^2.
\label{ss}
\ee
Using the identity
\be
\int {d^3p_2\over {2 E_2}} =
\int d^4p_2 \Theta (E_2) \delta(p_2^2 - m_{\nu}^2),
\label{dp4}
\ee
where $\Theta (E_2)$ is the step function and $m_{\nu}$ is the
neutrino mass. Putting eq.(\ref{dp4}) in 
eq.(\ref{ss}) and integrating over $p_2$ we obtain for energy 
radiated by neutrino in time T is
\be
S={T\over {16\pi^2 E_1}}\int {d^3k\over {2{\bf |p_1||k|}}}
\delta\Big ( {{(2E_1 w-w^2 +k^2)}\over {2 {\bf|p_1||k|}}} - cos\theta\Big )
|{\cal M}|^2.
\label{si}
\ee
The angle $\theta$ between the incoming neutrino and 
the emitted plasmon is obtained from the delta function in eq.(\ref{si}),
\be
cos\theta = {(2 E_1 w - w^2 + k^2)\over {2{\bf |p_1||k|}}}
={1\over {nv}}\Big (1 + {(n^2-1) w\over {2 E_1}}\Big ),
\ee
where $v ={|{\bf p_1}|\over E_1}$ is the neutrino velocity ($\simeq$ 1). Since
$-1 \le cos\theta \le 1$; which implies
\be
-{2 E_1\over (n-1)}\le w \le {2 E_1\over (n+1)}.
\ee
But definitely $-{2 E_1\over (n-1)}$ can not be the lower limit for the 
above Cerenkov process, as for $n >1$ this is a negative quantity and 
$w$ can not be negative. On the other hand from the dispersion relation 
for the longitudinal photon we obtain $w \ge 0.035 w_p$.
Thus the kinematically allowed region for the Cerenkov process is
\be
0.35 w_p\le w \le {2 E_1\over {(n+1)}}.
\ee
Evaluating $|{\cal M}|^2$ and simplifying the eq.(\ref{si}) we obtain
\be
S={T\over {8\pi^2\alpha}}G^2_F C_V^2 \int^{w_2}_{w_1}
{(n^2-1)^3\over n^4} w (1-{w\over E_1}) P_{l}^2 dw,
\ee
where $C_V= (2 sin^2\theta_W \pm {1\over 2})$ for $\nu_e,~\nu_{\mu}$ and
$w_1$ and $w_2$ are the lower and upper limits of $w$, $\theta_W$ is the weak
mixing angle and $sin^2\theta_W\simeq 0.233$.
As the plasmon emission is possible for a narrow range of the refractive
index $1 < n \le 1.0185$, we assume $n(w)\simeq n$ and take an average 
value of $n$ within the above range ($n=1.006$). 
Assuming the plasma frequency to be much smaller than the
incoming neutrino energy, $w_p << E_1$ we obtain
\be
S\simeq {T\over {8\pi^2\alpha}} G^2_F C^2_V P_{l}^2 (n-1)^3 E_1^2.
\ee
Thus the energy intensity of the emitted longitudinal photon by 
neutrino is proportional to the square of the incoming neutrino
energy.

\section{Supernova shock revival}
      
  Observation of neutrino events from supernova SN1987A  confirms that
neutrino emission is an efficient process
of cooling, hot, dense and collapsed stars. The production and propagation
of neutrinos are greatly influenced by the collective effects of the stellar
plasma. Adams, Ruderman and Woo\cite{adams} 
pointed out that plasmon decaying into
neutrino-antineutrino pair ($\gamma \rightarrow \nu{\bar\nu}$) would play
an important role in the stellar cooling process. Neutrinos are also
responsible for the delayed explosion of Type-II supernova.
Recent numerical calculations in more than one dimension shows that material
behind the stalled shock wave of the supernova can be heated efficiently
by neutrinos coming from the neutrinosphere and
eventually expel the outer mantle 
causing the supernova explosion\cite{arnet,burrows,akhmedov}. 
The neutrino properties get modified when propagate through the plasma
medium because of the weak interaction with the background particles.
Thus plasma process is of great importance in studying the astrophysical 
problems.
Here we consider the energy deposited by the Cerenkov 
process described above in the stalled shock wave of 
the supernova and compare with the
Bethe-Wilson mechanism of the shock revival\cite{bethe}.

  After the gravitational collapse of a massive star into neutron star, a
shock wave is formed and after traveling some distance (about 400 
Km \cite{fuller})
get stalled because most of the kinetic energy in the shock wave is used
to dissociate the nuclei. Bethe and Wilson in 1982 
showed\cite{bethe} that,
neutrinos from the hot inner core of the supernova are captured by the 
matter behind the shock through the process
$\nu_e + n\rightarrow p+e^-$ and ${\bar\nu_e} + p\rightarrow n+e^+$  and
deliver their energy. About 0.1\% of the total energy is sufficient to 
reheat the matter and eject the stalled shock. In BW mechanism 
rate of energy absorbed
by a gram of matter at a distance R is
\be
{\dot E_{BW}}=3\times 10^{18} L_{\nu 52} 
\Big ( {T^2_{\nu}\over R^2_7}\Big )~
{\tilde Y_N}~~ erg/g/sec,
\ee
where $L_{\nu 52}$ is the neutrino luminosity 
in units of $10^{52}$ erg/sec,
$R_7$ is the distance from the center in units of $10^7$ cm, $T_{\nu}= 5$
MeV is the temperature of the neutrino sphere and $\tilde Y_N \simeq 1$
is the total mean fraction of the nucleon. Here we neglect the contribution
due to electron and positron capture as they are correction to this
contribution. The total energy absorbed by the stalled shock wave,
which has a thickness $d$ and density $\rho$  is 
\be
{\dot E_{BW}} = 3\times 10^{18}~ L_{\nu 52} ~\Big ( {T^2_{\nu}\over R^2_7})
~~{\tilde Y_N}~~ erg/g/sec \times 4\pi\rho R^2 d.
\label{bw}
\ee
For the neutrino luminosity $L_{\nu_e} = 4\times 10^{52}~~erg/sec$ 
and the stalled
shock wave density $\rho\simeq 10^8~ g/cm^3$\cite{woosley} and $R =400$ Km,
the energy absorbed by the shock wave (assuming 100\% absorption) is
\be
{\dot E_{BW}}= 9.4\times 10^{43}~ d_{cm}~~ erg/sec. 
\ee
So 0.1\% of this  is
$9.4\times 10^{42}~d_{cm}~~erg/sec$, where $d_{cm}$ is units of cm.

In the Cerenkov process the total energy emitted by neutrinos
(we assume that all the radiated photons is absorbed 
by the medium) per unit time
is 
\bary
{\dot E_C} &=& S \times ~~(Neutrino~ flux)
\nonumber\\
&=& {d\over {8\pi^2\alpha}} G^2_F C^2_V P_{l}^2 (n-1)^3 E_1^2\times
\Big ({L_{\nu}\over E_1}\Big ).
\eary
Here we have replaced $T$ by the thickness of the medium $d$.
For average electron neutrino energy $E_1=12$ MeV we obtain 
${\dot E_C}~ =~ 1.8\times 10^{36}~d_{cm}~~erg/sec$.
Now comparing both the process (0.1\% of BW)
we have
\be
{{\dot E_{C}}\over {\dot E_{BW}}}\simeq 1.9\times10^{-7}.
\ee
This shows that the energy deposition by Cerenkov process is extremely
small.  Both the BW process and the Cerenkov process are weak processes,
so in principle the energy deposition in both the processes should not
differ too much, and even if there is difference, it should be compensated
by the electromagnetic coupling of the neutrino in the medium. 
But as the Cerenkov process is kinematically allowed
only for a very small range of the refractive index $n$ very close to
unity and the energy emitted is proportional to $(n-1)^3$; for 
refractive index $n\simeq 1.006$, $(n-1)^3$ is of order $10^{-7}$ 
thus reduces the contribution for Cerenkov process.

          It is a great pleasure to thank Dr. S. Mohanty for many
useful discussions.

\vfill
\eject

\bibliographystyle{plain}

\vfill
\eject
\end{document}